\documentclass[12pt]{article}
\pdfoutput=1
\usepackage[margin=1in]{geometry} 
\usepackage{natbib}

\usepackage{appendix,soul}

\usepackage{graphicx}
\usepackage{tabularx,booktabs,array,dcolumn}
    \newcolumntype{L}[1]{>{\raggedright\let\newline\\\arraybackslash\hspace{0pt}}m{#1}}
    \newcolumntype{C}[1]{>{\centering\let\newline\\\arraybackslash\hspace{0pt}}m{#1}}
    \newcolumntype{R}[1]{>{\raggedleft\let\newline\\\arraybackslash\hspace{0pt}}m{#1}}
    \newcolumntype{d}[1]{D{.}{.}{0}}
\usepackage{caption,subcaption}
\usepackage{chngpage}
\usepackage{eurosym}
\usepackage{float,rotating,placeins}
\usepackage{color,xcolor}
\usepackage{amsmath,amsthm,amssymb,bbold,amsfonts,mathtools}
    \allowdisplaybreaks
\usepackage{setspace}
\usepackage[bottom]{footmisc}
\usepackage{accents} 	
\usepackage[pdftex,
			pdfauthor={},
            pdftitle={},
            pdfsubject={},
            pdfkeywords={},
			hyperfootnotes=false,bookmarksopen=true]{hyperref}
\hypersetup{colorlinks=true,allcolors=blue}

\DeclareMathOperator*{\E}{\mathbb{E}}

\DeclareMathOperator*{\Cov}{\mathrm{Cov}}

\newcommand{\fsz}{\footnotesize}

\newcommand{\mcl}{\multicolumn}

\title{\textbf{Income Shocks}\\\Large \textbf{and their Transmission into Consumption\thanks{Crawley: Federal Reserve Board; email: \href{mailto:edmund.s.crawley@frb.gov}{edmund.s.crawley@frb.gov}. Theloudis: Department of Econometrics \& OR, Tilburg University; email: \href{mailto:a.theloudis@gmail.com}{a.theloudis@gmail.com}. Viewpoints and conclusions stated in this article are the responsibility of the authors alone and do not necessarily reflect the viewpoints of the Federal Reserve Board. This review will appear in the Encyclopedia of Consumption.}}}

\author{ 
Edmund Crawley
\and
Alexandros Theloudis
}
\date{April 18, 2024}

\begin{document}
\maketitle


Measuring how household consumption responds to income shocks is important for understanding how families cope with adverse events, for designing government insurance or other income support policies, and for understanding the transmission of business cycles and monetary policy. It is also important for evaluating the effects of fiscal or labor market reforms on consumer welfare, and for how these reforms may impact the macroeconomy given that consumption is a large share of GDP. 

This article reviews the economics literature of, primarily, the last 20 years, that studies the link between income shocks and consumption fluctuations at the household level. We identify three broad approaches through which researchers estimate the consumption response to income shocks: 1.) structural methods in which a fully or partially specified model helps identify the consumption response to income shocks from the data; 2.) natural experiments in which the consumption response of one group who receives an income shock is compared to another group who does not; 3.) elicitation surveys in which consumers are asked how they expect to react to various hypothetical events. None of these approaches is exclusive to a single field within economics; studies that use any of these methods are ordinarily classified, depending on their specific focus, in macroeconomics, labor economics, public finance -- to name only a few fields.  

Our aim in this short article is to survey this increasingly busy literature and provide an accessible summary of the various estimates of the consumption response to income shocks. We concentrate on the similarities and differences between the various studies, in particular with respect to the method, data, consumption notion, and type of income shock analyzed, thus also with respect to the type of consumption response each work identifies. Our focus is on responses to shocks, i.e., \emph{unanticipated} income changes; \citet{JappelliPistaferri2010ConsumptionResponse} review the earlier evidence on responses to \emph{anticipated} income changes.

The survey proceeds as follows. Section \ref{Section::TheoreticalBackground} introduces a brief theoretical framework that helps fix ideas for the subsequent discussion. The next sections are devoted to the different approaches to the estimation of the consumption response. Section \ref{Section::StructuralMethods} surveys the studies that employ structural methods, section \ref{Section::NaturalExperiments} reviews the evidence from natural experiments, and section \ref{Section::ElicitationSurveys} focuses on the elicitation surveys. Section \ref{Section::Conclusion} concludes.

\section{Theoretical Background}\label{Section::TheoreticalBackground}

A household $i$ chooses consumption $C_{it}$, savings $A_{it+1}$, and perhaps other behaviors captured in $L_{it}$ (possibly a vector), e.g., labor supply, to maximize its expected lifetime utility
\begin{equation}\label{Eq::Household_Problem}
    \max_{\{C_{it},A_{it+1},(L_{it})\}_{t=0}^{T}} {\E}_{0} \sum_{t=0}^{T} \beta^{t} U_i(C_{it};L_{it}),
\end{equation}
for which we assume separability over time and geometric discounting.\footnote{Strictly, the household chooses a plan for $C_{it}$, $A_{it+1}$, and $L_{it}$ contingent on future states of the world, including, e.g., future states of income. We do not explicitly show the contingent states to ease the notation.} Expectations about future states of the world are captured by ${\E}_{0}$. The utility function depends on consumption and on $L_{it}$, which may be a choice variable (e.g., endogenous labor supply) or taken as given (e.g., exogenous labor supply). Utility is subscripted by $i$ to reflect preference heterogeneity across households. We have employed a finite-horizon setting with $T$ as its terminal period, which is natural given our focus on households.\footnote{The extension to infinite horizon is trivial and mostly inconsequential \citep{JappelliPistaferri2010ConsumptionResponse}.}

The problem is subject to the sequential budget constraint
\begin{equation}\label{Eq::Budget_Constraint}
	(1+r)A_{it} + T_i(Y_{it}^g; L_{it}) = C_{it} + A_{it+1}, 
\end{equation}
which links resources over time under the assumption that consumers can borrow and save at an interest rate $r$.\footnote{The extension to multiple or risky assets is also straightforward.} $T_i$ maps before taxes/transfers gross income $Y_{it}^g$ to disposable household income $Y_{it}^d$. $Y_{it}^g$ may be a vector; for instance, with two earners in the household, $Y_{it}^g=(Y_{it}^{e_1},Y_{it}^{e_2})^\prime$, where $Y^{e_j}$ indicates earnings of member $j$. $T_i$ may depend on choices $L_{it}$; for example, with endogenous labor supply, the primitive source of gross income is the hourly wage, i.e., $Y_{it}^g=(Y_{it}^{w_1},Y_{it}^{w_2})^\prime$, where $Y^{w_j}$ indicates the wage of member $j$. $T_i$ is subscripted by $i$ to reflect heterogeneity in taxes, welfare benefits, contingent transfers, or external sources of income (what is often called external insurance). There may be a borrowing constraint in some periods such that $A_{it} \geq B_{it}$, where $B_{it}$ denotes the applicable borrowing limit. Finally, there is a terminal condition on $A_{iT+1}$, which reflects that households run down their assets before death or bequeath their offspring.

There are many alternative specifications for the process that governs income; a general formulation is
\begin{equation}\label{Eq::IncomeProcess}
	Y_{it}^k = f_i^k(X_{it},Y_{it-1}^k,v_{it}^k,u_{it}^k,\dots),
\end{equation}
where $k=\{d,g,e_1,e_2,w_1,w_2, \dots\}$ indicates the type of income considered, i.e., disposable, gross, etc. $f_i^k$ reflects the precise process, which allows for heterogeneity across households and depends on observables $X_{it}$ (e.g., age, time, education), past income $Y_{it-1}^k$, and idiosyncratic shocks $v_{it}^k$ and $u_{it}^k$ to \emph{log} income, such as permanent and transitory shocks. It may also depend on older income, other shocks including aggregate ones, and past shocks, depending on the specific process that is being considered. \citet{MeghirPistaferri2011Handbook} offer a review of the vast income dynamics literature and discuss popular cases.\footnote{There are multiple generalizations of \eqref{Eq::IncomeProcess}. For example, one may allow $f_i^k$ to depend on time (age), thus enable the impact of shocks to be time-varying. We view \eqref{Eq::IncomeProcess} as a simple organizing device rather than as an exhaustive representation of every possible income process.}

Solving \eqref{Eq::Household_Problem} subject to \eqref{Eq::Budget_Constraint}, the borrowing constraint, and the terminal condition, yields a consumption policy rule whose exact formulation depends on the preference specification, the income process \eqref{Eq::IncomeProcess}, the tightness of the borrowing constraint, and the market environment in which the household operates (for example, the extent to which it has access to contingent transfers or external insurance). A general formulation for the consumption rule is 
\begin{equation}\label{Eq::ConsumptionRule}
	C_{it} = g_i(A_{it},X_{it},Y_{it-1}^k,v_{it}^k,u_{it}^k,\dots),
\end{equation}
which allows for heterogeneity across households (reflecting, among other things, heterogeneity in $U_i$) and generally depends on assets and the various components of income. Policy rule \eqref{Eq::ConsumptionRule} subsumes several popular settings in the literature, and \citet{JappelliPistaferri2010ConsumptionResponse} offer specific examples. It is nonetheless non exhaustive of all possible settings. For instance, durable goods necessitate accounting for their stock and possible adjustment costs. Here we simply see \eqref{Eq::ConsumptionRule} as a general organizational device rather than the solution to any given model. 

Interest lies in how and by how much the income shocks $v_{it}^k$, $u_{it}^k$, etc, impact consumption. To measure this, the literature focuses on two main parameters, the marginal propensity to consume (MPC), broadly defined as the derivative of consumption with respect to income -- i.e., $dC_{it} / dY_{it}^k$,  and the pass-through rate, broadly defined as the derivative of consumption growth with respect to the shock -- i.e., $d\Delta \ln C_{it} / dv_{it}^k$. We now turn to the three broad approaches to estimating these parameters.

\section{Structural Methods}\label{Section::StructuralMethods}

\noindent \textbf{Early papers.} A series of influential papers in the 1980s-90s test the predictions of the permanent income hypothesis and the complete-markets model, the then benchmark models in the literature. This is done through forming appropriate hypotheses on the link between consumption and income fluctuations that emanate from these models. While this early work does not strictly \emph{measure} MPCs or pass-through rates, it provides motivating evidence for the subsequent work that explicitly measures the consumption response to shocks.

\citet{Hall_Mishkin1982}, one of the first studies, investigate the sensitivity of food consumption to income using micro data from the Panel Study of Income Dynamics (PSID). Consumption varies more closely with permanent than with transitory shocks, which is an implication of the permanent income hypothesis in which households smooth consumption through self-insurance (saving, borrowing). Yet, the sensitivity of consumption to transitory income is much stronger than theory predicts, which leads to rejection of the model.\footnote{Other early papers that test the permanent income hypothesis are \citet{Hall1978StochasticImplicationsLifeCyclePermanentIncomeHypothesis}, who does not reject it, \citet{Sargent1978RationalExpectations}, who rejects it, and \citet{Flavin1981TheAdjustmentConsumption}, who also rejects it. They all use time series data.}

The permanent income hypothesis postulates that inequality in consumption grows over the lifecycle. This is the motivating observation for \citet{Deaton_Paxson1994} who, using micro data from multiple countries, confirm that the variance of consumption (and income) grows with age. While they cannot reject the permanent income hypothesis, they admit that the evidence is also consistent with other models of intertemporal choice, such as models that permit some external insurance to idiosyncratic income shocks.

On the opposite end of theory, \citet{Cochrane1991_TestConsumptionInsurance} tests for full insurance by assessing the sensitivity of consumption to income growth and to events such as illness and job loss. Under complete markets, consumers have access to contingent transfers, so household consumption growth should be unrelated to \emph{idiosyncratic} events. Focusing on food consumption in the PSID, the hypothesis is rejected following long illness or job loss, but not rejected in response to short unemployment spells, thus providing early evidence for partial insurance.\footnote{\citet{AltugMiller1990HouseholdChoicesEquilibrium} model a complete-markets environment and allow for non-separability with labor supply; using food consumption in the PSID, they cannot reject full insurance to wage fluctuations.}

\citet{AttanasioDavis1996_RelativeWageMovements} test for full insurance across birth cohorts and education groups, accounting for consumption-work complementarity and common demographics driving consumption and wages. Under complete markets and absent of aggregate shocks, consumption growth should not co-move with wage growth. Drawing synthetic panels from the Consumer Expenditure (CEX) and Current Population (CPS) surveys, they sharply reject full insurance to low-frequency wage shifts. In a similar model, \citet{Hayashi_Altonji_Kotlikoff1996} test for full insurance across and within extended families, which they reject based on the strong correlation between wage and food consumption growth in the PSID. Using similar data in a simpler setting, \citet{AltonjiHayashiKotlikoff1992ExtendedFamilyAltruisticallyLinked} also reject full insurance. By contrast, \citet{Mace1991FullInsuranceAggregateUncertainty} and, in particular, \citet{Townsend1994RiskInsuranceVillageIndia}, who studies village insurance in India, find mixed evidence.

In sum, these early works frequently reject the benchmark models of permanent income (self insurance) and complete-markets (full insurance). Yet, the data consistently reveal that households have access to \emph{some} insurance to income shocks.\footnote{Among the first studies to quantify the extent of consumption insurance, \citet{Gruber1997UnemploymentInsurance} measures how UI reduces the fall in consumption upon unemployment.} The literature in the busy 2000s-10s attempts to measure the degree of insurance and identify its sources.

\

\noindent \textbf{Covariance restrictions.} \citet{Blundell_Pistaferri_Preston2008}, abbreviated as BPP, introduce the seminal methodology to measure the consumption response to income shocks. Their idea is that the extent to which consumption growth varies with income growth, the latter being driven by various income shocks, reflects the \emph{degree} of transmission of those shocks into consumption.\footnote{In a predecessor paper, \citet{Blundell_Preston1998} assume that that the rise in consumption inequality observed by \citet{Deaton_Paxson1994} is driven by permanent but not transitory shocks, to which consumers can fully self-insure. This allows them to use income and consumption moments to identify the variances of permanent and transitory income shocks. \citet{AttanasioEtAl2002FromEarningsToConsumption} extend this to a setting of two earners with separate income streams, while \citet{PrimicerivanRens2009HeterogeneousProfiles} extend it to heterogeneous income processes.} This is motivated by a consumption process that is log-linear in income shocks, namely 
\begin{equation}\label{Eq::ConsumptionProcess_BPP}
    \Delta c_{it} = \xi_{it} + \phi_t v_{it}^k + \psi_t u_{it}^k,
\end{equation}
where $\Delta c_{it}$ is consumption growth $\Delta \ln C_{it}$ net of observables $X_{it}$.\footnote{We use $\Delta$ to denote the first difference operator; so $\Delta X_{t} = X_{t} - X_{t-1}$.} $v_{it}^k$ and $u_{it}^k$ are, respectively, a permanent and a transitory shock to disposable household income (so $k=d$ here), $\phi_t$ and $\psi_t$ are their transmission parameters, and $\xi_{it}$ is a preference shock unrelated to income. If income follows the canonical permanent-transitory process, namely $\Delta y_{it}^k = v_{it}^k + \Delta u_{it}^k$ where $\Delta y_{it}^k$ is income growth $\Delta \ln Y_{it}^k$ net of $X_{it}$, the transmission parameters are identified through ${\Cov}(\Delta c_{it}, Z_{it})/{\Cov}(\Delta y_{it}^k, Z_{it})$, namely a regression of $\Delta c_{it}$ on $\Delta y_{it}^k$ using appropriate instruments $Z$ for income. In the case of permanent shocks, permanent income $Z_{it} = \sum_{\tau=-1}^{1}\Delta y_{it+\tau}^k$ nets $\Delta y_{it}^k$ from the transitory shock at $t$, so its covariance with $\Delta c_{it}$ identifies $\phi_t$. In the case of transitory shocks, future income $Z_{it} = \Delta y_{it+1}^k$ shifts $\Delta y_{it}^k$ due to mean reversion of the transitory shock, so its covariance with $\Delta c_{it}$ identifies $\psi_t$. It is precisely these covariance restrictions that this strand of literature has taken its name from. 

The consumption equation \eqref{Eq::ConsumptionProcess_BPP} can be obtained through a log-linearization of the policy rule \eqref{Eq::ConsumptionRule} in a lifecycle permanent income model with CRRA utility, a permanent-transitory income process, and slack borrowing constraints. In this case, $\phi_t$ reflects the share of consumption that is funded by future labor income (as opposed to assets whose value remains unchanged by the shock to income) and, depending on the measure of $Y_{it}^k$, features of the tax/benefits system; $\psi_t$ is similar up to an annuitization factor for the household's remaining horizon. Yet, \eqref{Eq::ConsumptionProcess_BPP} is also consistent with other settings in which the transmission of shocks depends on their persistence, e.g., environments with moral hazard \citep{AttanasioPavoni2011PrivateInformationModels} or external insurance \citep{Blundell_Pistaferri_Saporta-Eksten2016}. The linear insurance equation \eqref{Eq::ConsumptionProcess_BPP} is, therefore, the \emph{reduced form} of multiple environments that may differ in their insurance content. As such, $\phi_t$ and $\psi_t$ are called \emph{partial insurance} parameters; they measure the \emph{overall} pass-through of (or, on the flip side, insurance to) permanent and transitory income shocks, regardless of the precise mechanisms that give grounds for such insurance.

Empirically, BPP focus on nondurable consumption and disposable household income in 1980-92. While the PSID is ideal for its income data, its consumption data, including mostly food items, is limited until 1999; BPP thus impute consumption from the CEX into the PSID. They estimate $\phi_t = 0.64$ (a 10\% permanent income cut reduces consumption by only 6.4\% -- households are thus partially insured to permanent shocks) and $\psi_t = 0.05$ (statistically not different from zero, so consumption is fully insured against transitory shocks).

BPP sparked multiple extensions. \citet{BlundellPrestonLow2013decomposing} derive \eqref{Eq::ConsumptionProcess_BPP} under an autoregressive income process and general preferences. They assess the approximation vis-\`a-vis the true policy rule and show that it performs well when liquidity constraints do not bind. \citet{Blundell_Pistaferri_Saporta-Eksten2016} model endogenous labor supply to measure the consumption response to \emph{wage} shocks. Using income and consumption data in the PSID after 1999, they estimate $\phi_t=0.32/0.19$ (male/female wages) and find little external insurance when family labor supply, assets, and the tax/benefits system are accounted for.\footnote{\citet{Hyslop2001IntrafamilyEarnings} uses covariance restrictions to measure the pass-through of wage shocks into household earnings. \citet{JessenKonig2023HoursWageRisk} is related: using hours and earnings data in the PSID over 1970-1997, they find that wage and taste shocks have a comparable contribution to the total variance of earnings.} \citet{Theloudis2017_PhDThesis} extends this to an intra-household bargaining (collective) setting, in which lack of commitment to lifetime marriage limits the insurance role of family labor supply. \citet{Chopra2023InsuranceCyclicality} argues that the insurance role of family labor supply increases during recessions.

Most of these works permit limited heterogeneity. \citet{Arellano_Blundell_Bonhomme2017}, ABB in short, relax linearity in the income process and let shocks feature \emph{nonlinear} persistence depending on their sign and size. The consumption response to shocks varies flexibly with their level and past history. Using a quantile-based estimation method and PSID data over 1999-2011, they measure the consumption response to log income, a type of MPC, at 0.2-0.4 on average (specification with household heterogeneity), though the response varies over the distribution of shocks. Although allowing for heterogeneity makes interpreting their parameter similar to a pass-through rate, the response is markedly different from BPP. This may be due to ABB's flexible income process or the new consumption data in the PSID.\footnote{ABB is not a covariance method, but their framework falls firmly in this category because they measure the \emph{overall} impact of shocks, as in BPP. \citet{Arellano_Blundell_Bonhomme_Light2021} advance this to unbalanced panels and flexible heterogeneity. They measure the consumption response to log income at 0.2 on average.}

\citet{GhoshTheloudis2023TailIncomeRisk} relax linearity in the consumption process by writing $\Delta c_{it}$ as a quadratic polynomial in income shocks. This stems from a second-order approximation to the policy rule \eqref{Eq::ConsumptionRule} in a model similar to BPP. The pass-through of shocks now depends on their sign and size, so this method allows small vs. large, or good vs. bad shocks, to have an \emph{asymmetric} impact on consumption. Identification requires second, third, and fourth moments of income and second moments of consumption, in contrast to ABB who require observing their entire distribution. Using PSID data after 1999, they estimate the pass-through of the \emph{average} permanent shock at 0.13; bad or large permanent shocks have a much bigger impact on consumption, and their pass-through increases with their severity.\footnote{The average pass-through is close to ABB but markedly different from BPP. The discrepancy is due to the consumption imputation in BPP and the biennial (vs. annual) frequency of the modern data.}

\citet{AlanBrowningEjrnaes2018IncomeAndConsumption} estimate the extent of insurance to income shocks allowing for flexible joint heterogeneity in the consumption and income processes. They use PSID data over 1968-2009 and find pervasive cross-household heterogeneity in the pass-through, ranging from 0.05 to 0.69. \citet{Theloudis2021} allows for unobserved preference heterogeneity in a model with family labor supply. He explores higher-order moments of income and consumption to identify the contribution of heterogeneity to consumption inequality.

A consistent empirical finding in this literature is that, on average, consumption is fully insured against transitory shocks. Several papers challenge this result. \citet{Crawley2020LostTimeAggregation} argues that BPP neglect time aggregation in the PSID. Time aggregation occurs when income is observed less frequently (annually) than the underlying true data (e.g., monthly payments); income growth is then \emph{mechanically} positively correlated, which changes the covariance restrictions used to identify the variance of shocks and their transmission into consumption. \citet{Crawley2020LostTimeAggregation} addresses this point and estimates the pass-through of shocks at $\phi_t = 0.34$ and $\psi_t = 0.24$ (statistically significant).

\citet{Commault2022} extends the linear insurance model \eqref{Eq::ConsumptionProcess_BPP} to allow consumption to respond to \emph{past} transitory shocks. There are several underlying structures that justify this. The exact solution to $\Delta c_{it}$ in a lifecycle permanent income model includes higher-order terms that depend on past variables (e.g., wealth); models with limited commitment is another example. \citet{Commault2022} proposes a new estimator for $\psi_t$, one that selects as instrument the only future $\Delta y_{it+\kappa+1}^k$ that is uncorrelated with past shocks. She estimates $\psi_t=0.6$ (MPC at 0.32), which helps bridge the discrepancy in $\psi_t$ between studies that employ covariance restrictions and those that rely on natural experiments.\footnote{As we review subsequently, natural experiments often imply a larger $\psi_t$ than typically estimated through covariance restrictions. It is unclear, however, if survey data such as the PSID, on which the latter studies rely, reflect larger transitory shocks that are typical of experiments. Studies that allow the pass-through to depend on the size of the shock typically find that larger shocks have larger pass-through, thus also helping bridge the gap in estimates between natural experiments and covariance restrictions.}

\citet{crawley2023consumption} address neglected time aggregation \emph{and} allow consumption to depend on past transitory shocks. They estimate heterogeneous $\phi_t$ and $\psi_t$ over well-defined subpopulations using Danish administrative data, and relate their estimates to liquid wealth and other household balance-sheet characteristics.

\citet{Hryshko_Manovskii2022} identify households in the PSID with vastly different degrees of insurance. They show that the \emph{sons} of families originally surveyed by the PSID in 1968 exhibit almost no insurance, while the \emph{daughters} have substantial partial insurance. They do a thorough job explaining this discrepancy by differential income persistence across the two groups, which is further explained by differential attrition from the survey. 

\

\noindent \textbf{Fully specified models.} The studies employing covariance restrictions do not \emph{fully} specify preferences, expectations, and the budget set, so they take no stance on the exact mechanisms that give rise to partial insurance. Due to their semi-parametric nature, these works are of limited use for policy counterfactuals. Another approach is to fully specify the channels through which consumers smooth income shocks, and take such models to the data.

\citet{AttanasioLowSanchez2005FemaleInsurance} quantifies the insurance role of female labor supply (measured in terms of welfare costs of income uncertainty) through a structural model of consumption and labor supply with earnings risk. \citet{Krueger_Perri2006} show that a model with a complete set of Arrow securities but limited enforceability of contracts reproduces income and consumption inequality in the U.S., suggesting that households possess more insurance than self-insurance. \citet{StoreslettenTelmerYaron2004ConsumptionRiskSharingLifeCycle} show that a lifecycle permanent income model can also produce empirically consistent income and consumption inequality if the tax/benefits system and the aggregate level of wealth are accounted for. \citet{HeathcoteStoreslettenViolante2008WelfareAnalysisLaborMarketRisk} calibrate a lifecycle model of consumption/labor supply with partial insurance to wage shocks to measure the welfare costs of risk and market incompleteness. Partial insurance is fixed by the authors: permanent shocks are uninsured, while transitory shocks are fully insured. \citet{LowMeghirPistaferri2010WageEmploymentRisk} calibrate a lifecycle model of consumption, labor supply, and job mobility with income and employment risk. Self-insurance aside, the model allows for three channels of partial insurance: unemployment benefits, disability insurance, and food stamps.

\begin{table}[t!]
\begin{center}
\caption{Summary of Estimates from Structural Methods}\label{Table::SummaryStructural}
\resizebox{\textwidth}{!}{%
\begin{tabular}{l C{1.4cm} C{1.5cm} C{1.4cm} C{1.6cm} C{1cm} C{1cm} L{3.1cm}}
\toprule
  						                                    & \mcl{2}{c}{\textbf{Pass-through}}   & \mcl{2}{c}{\textbf{MPC}}          & \mcl{2}{c}{Variables}&              \\
Study                                                       & perm.         	& trans.          & perm.           & trans.          & $Y^k_{it}$ & $C_{it}$& Data         \\
\midrule
\citet{AlanBrowningEjrnaes2018IncomeAndConsumption}         & \mcl{2}{c}{0.05$-$0.69}             &                 &                 & thy     & food       & P 1999-2009             \\
\citet{Arellano_Blundell_Bonhomme2017}\textsuperscript{a}   &    				& 	              & 0.2$-$0.4       & -0.4$-$0.2      & thy     & nde        & P 1999-2009             \\
\citet{Arellano_Blundell_Bonhomme_Light2021}\textsuperscript{b}&                &                 & 0.33            &                 & dhy     & nde        & P 2005-17               \\
\citet{Blundell_Pistaferri_Preston2008} 			        & 0.64              & 0.05            &                 &                 & dhy     & nde        & P \& C 1980-92          \\
\citet{Blundell_Pistaferri_Saporta-Eksten2016}		        & 0.32   	    	& -0.14       	  &                 &                 & mhw    	& nde        & P 1999-2009             \\
															& 0.19    	    	& -0.04       	  &                 &                 & fhw    	& nde        & P 1999-2009             \\
\citet{BlundellPistaSaporta2018ChildrenTimeAllocation}      & 0.39              & 0.12            &                 &                 & mhw     & nde        & P 1999-2015, and        \\
															& 0.35              & 0.13            &                 &                 & fhw     & nde        & ~~C \& A 2003-15         \\
\citet{Busch_Ludwig2021}\textsuperscript{c}                 & 0.40              & 0.05            & 0.38            & 0.05            & dhy     & n/a        & P 1977-2012             \\
\citet{Chopra2023InsuranceCyclicality}                      & 0.29 r            & -0.18 r         & 0.19 r          &                 & mhw     & nde        & P 1977-2016             \\
															& 0.31 x            & -0.26 x         & 0.12 x          &                 & mhw     & nde        & P 1977-2016             \\
\citet{Commault2022}                                        &                   & 0.6             &                 & 0.32            & dhy     & nde        & P \& C 1980-92          \\
\citet{Crawley2020LostTimeAggregation}                      & 0.34              & 0.24            &                 &                 & dhy     & nde        & P \& C 1980-92          \\
\citet{crawley2023consumption}                              &                   &                 & 0.64            & 0.64           & dhy     & te         & D 2003-15               \\
\citet{DeNardi_Fella_PazPardo2019}                          & 0.54              & 0.12            &                 &                 & dhy     & nde        & P 1968-92, and          \\
															&              		&                 &                 &                 &      	&         	 & ~~C 1980-2007           \\
\citet{GhoshTheloudis2023TailIncomeRisk}\textsuperscript{c} & 0.13              & -0.00           &                 &                 & dhy     & nde        & P 1999-2019             \\
\citet{Guvenen_Smith2014}                                   & \mcl{2}{c}{0.45}                	  &                 &                 & dhy     & nde        & P \& C 1968-92          \\
\citet{GuvenenMaderaOzkan2023_ConsumptionResponseTailShocks}\textsuperscript{d}& 0.38 & 0.11      & 0.4             & 0.05            & dhy     & nde        & external estims 		   \\
\citet{Heathcote_Storesletten_Violante2014}                 & 0.39              &                 &                 &                 & mhw     & nde        &  P 1968-2007, and       \\
															&              		&                 &                 &                 &      	&         	 & ~~C 1980-2006           \\
\citet{Hryshko_Manovskii2022}                               & 0.87 sn           & 0.07 sn         &                 &                 & dhy     & nde        & P \& C 1980-92          \\
															& 0.46 dg           & 0.12 dg         &                 &                 & dhy     & nde        & P \& C 1980-92          \\
\citet{JessenKonig2023HoursWageRisk}    			        & 0.62              &                 &                 &                 & mhw     & n/a        & P 1970-1997             \\
\citet{Kaplan_Violante2010}                                 & 0.78              & 0.06            &                 &                 & dhy     & nde        & P 1980-92, SCF          \\
\citet{LowMeghirPistaferri2010WageEmploymentRisk}\textsuperscript{e}& 0.56      &                 &                 &                 & mhw     & n/a        & P 1988-96, SIPP         \\
\citet{Madera2019_ConsumptionResponseTailShocks}\textsuperscript{d}& 0.50       & 0.10            &                 &                 & dhy     & te         & P 1999-2015             \\
\citet{Theloudis2021}                                       & 0.45              & -0.03           &                 &                 & mhw     & nde        & P 1999-2011             \\
															& 0.27              & -0.05           &                 &                 & fhw     & nde        & P 1999-2011             \\
\citet{WuKrueger2021_ConsumptionInsurance}                  & 0.35              & 0.01            &                 &                 & mhw     & nde        &  P 1999-2009             \\
															& 0.18              & 0.01            &                 &                 & fhw     & nde        &  P 1999-2009             \\
\bottomrule
\end{tabular}}
\caption*{\fsz\emph{Legend:} \textbf{dg}: daughters; \textbf{dhy}: disposable household income; \textbf{fhw}: female hourly wage; \textbf{mhw}: male hourly wage; \textbf{nde}: non-durable expenditure; \textbf{n/a}: not applicable; \textbf{r}: recession; \textbf{sn}: sons; \textbf{te}: total expenditure; \textbf{thy}: total household income; \textbf{x}: expansion; \textbf{A}: American Time Use Survey; \textbf{C}: Consumer Expenditure Survey; \textbf{D}: Danish registry data; \textbf{P}: Panel Study of Income Dynamics; \textbf{SCF}: Survey of Consumer Finances; \textbf{SIPP}: Survey of Income \& Program Participation.\\
\textsuperscript{~~a}Results with unobserved household heterogeneity, figures S21 and S24.\\ 
\textsuperscript{~~b}Results with filtering and unobserved household heterogeneity.\\
\textsuperscript{~~c}Results for average/medium shock.\\ 
\textsuperscript{~~d}Results at age 40.\\ 
\textsuperscript{~~e}Pass-through of income following an unemployment shock.}
\end{center}
\end{table}

These earlier quantitative models measure the welfare implications of risk (or of certain insurance mechanisms) but not the degree of consumption insurance per se.\footnote{An exception is \citet{LowMeghirPistaferri2010WageEmploymentRisk} who report the consumption response to an unemployment shock.} \citet{Kaplan_Violante2010} explicitly measure the degree of insurance in a calibrated lifecycle permanent income model with income risk. This is the model whose consumption rule BPP log-linearize. While they find almost full insurance to transitory shocks, as in BPP, they estimate $\phi_t=0.78$, larger than BPP's estimate. Households in the model possess \emph{less} insurance to permanent shocks than in the data, highlighting that \emph{self-insurance}, the only mechanism in the model, is not enough to generate the excess consumption smoothing we observe empirically. By contrast, \citet{WuKrueger2021_ConsumptionInsurance} calibrate a lifecycle permanent income model with endogenous labor supply and find that it matches the empirical pass-through rates of male and female \emph{wage} shocks in \citet{Blundell_Pistaferri_Saporta-Eksten2016}. Family, and mainly \emph{female}, labor supply is a crucial insurance mechanism that previous studies had neglected.

\citet{Guvenen_Smith2014} estimate a lifecycle consumption–savings model with self-insurance, external insurance, and learning over stochastic income. They use the joint dynamics of earnings and consumption in the PSID and CEX to quantify earnings risk and the extent of insurance to it. About half of the earnings shock (including permanent and transitory elements) is smoothed. \citet{Heathcote_Storesletten_Violante2014} estimate a general equilibrium consumption/labor supply model with partial insurance to wage shocks. They model self-insurance, labor supply, a tax/benefits system, and external insurance. In their benchmark, they use PSID earnings and hours data alone and find that 39\% of permanent wage shocks pass through into consumption. \citet{BlundellPistaSaporta2018ChildrenTimeAllocation} present a structural lifecycle consumption–savings model with labor supply and childcare; in this setting, childcare responds endogenously to wage shocks and acts as an additional insurance mechanism.

The previous studies assume income risk is Guassian. \citet{Guvenen_Karahan_Ozkan_Song2021} and other authors establish that the distribution of income shocks exhibits substantial left skewness and excess kurtosis. This implies that there are far more people in the data who experience small unimportant or extreme negative shocks than people who experience moderate or extreme positive ones. A newer literature attempts to measure the pass-through of income shocks accounting for these higher-order features of income dynamics. 

\citet{DeNardi_Fella_PazPardo2019} estimate a lifecycle model of consumption/savings with non-Gaussian income risk. They show that tail income risk increases the degree of partial insurance to permanent shocks due to stronger precautionary motives. \citet{Busch_Ludwig2021} estimate a similar model explicitly targeting income skewness and kurtosis. They distinguish between good and bad shocks and find that the latter are worse insured than the former. In a related model, \citet{Madera2019_ConsumptionResponseTailShocks} studies the differential response of durable and nondurable consumption to tail earnings shocks. Durable consumption responds more strongly to tail shocks than nondurable consumption does. In a calibrated lifecycle consumption-savings model, \citet{GuvenenMaderaOzkan2023_ConsumptionResponseTailShocks} establish that non-Gaussian earnings risk implies large welfare losses and commands a strong consumption response. They confirm that the benchmark method in BPP understates the true consumption response to such shocks, as derived analytically by \citet{GhoshTheloudis2023TailIncomeRisk} in the presence of tail income risk.

\section{Natural Experiments}\label{Section::NaturalExperiments}
At the other end of the spectrum to structural models are reduced-form studies of natural experiments. These studies usually compare one group of individuals or households who have received a shock to their income, such as a stimulus check, with a group that has not. In contrast to studies of structural models, natural experiment studies tend to focus on estimating a marginal propensity to consume (MPC) as opposed to a pass through parameter. This is partly because the identified shock is not often proportional to income and indeed the researchers rarely know what household income is.

There are two main challenges for a researcher estimating MPCs using natural experiments. The first is to find a suitable natural experiment and the second is to find high-quality data on spending or consumption at the individual or household level. The literature has boomed recently as more and more data sources have become available to researchers. Some more recent papers are able to speak to MPC heterogeneity across household demographics and asset holdings, as well as shock size and sign.

Overall, the evidence from these natural experiments points to larger consumption responses to income shocks than from the structural modeling literature. However, the range of estimates is wide and there is still no consensus in the profession. \citet{havranek2020ruleofthumb} examine 144 studies of excess sensitivity and find evidence of publication bias that suggests MPCs may be smaller than implied by a survey of this literature such as this one.

\citet{parker2006rebates} was one of the first papers to study a convincing natural experiment. The authors look at household spending following the 2001 tax rebates in the United States. The distribution of these rebates, typically \$300 or \$600 in size, were staggered according to the second-to-last digit of recipients' social security number---an effectively random assignment. Using the CEX, the authors find that households spent between 20 and 40 percent of their rebates on non-durable goods during the three-month period in which they received their rebates, with some evidence that households with low liquid wealth or low income had larger MPCs. \citet{parker2013stimulus} use a similar research design on the 2008 stimulus checks in the United States and found a slightly smaller consumption response for nondurable goods, but a large total response---including durables---of 50 to 90 percent in the three-month period in which the check arrived. 

These estimates of the spending response to U.S. stimulus policies are highly cited and often contested. For example, \citet{MisraSurico2014_EvidenceFiscalStimulus} use quantile regression techniques on the same two experiments and find MPC estimates that tend to be smaller and more accurate than those from a homogeneous model. \citet{orchard2023counterfactuals}, while pointing to the difficulty of reconciling high MPC estimates with the macro evidence, also find lower MPC estimates when updating the results from \citet{parker2013stimulus} with new insights into the difference-in-differences methodology used. Pointing in the other direction is evidence that households significantly under report their spending on goods and non-housing services in the CEX---see \citet{sabelhaus2014CEXrepresentative}. Overall, these two episodes of government fiscal stimulus have been the subject of much research, partly because the random assignment provides an excellent identification scheme and partly because the effectiveness of such stimulus policies is a question of vital importance in its own right. 

\citet{shapiro2009taxrebates} and \citet{sahm2010household} study the 2008 tax rebates using questions on the Michigan Survey of Consumers. They find that around 20\% of respondents say that they will ``mostly spend'' the rebate when asked whether they would use the rebate to mostly spend, save, or pay off debt. The authors combine this with other evidence to suggest an implied MPC of around one third. Interestingly, \citet{sahm2012check}---who also use the Michigan Survey of Consumers---look at differences in consumption responses between a one-time payment relative to flow of payments from reduced tax withholding. They find that the one-time payment may be about twice as effective at inducing spending, highlighting some of the difficultly in pinning down consumer behavior.

During the pandemic, three Economic Impact Payments (EIPs) were distributed to households and their effects have been widely studied; \citet{falcettoni2021acts} and \citet{gelman2022lessons} provide more thorough reviews of the related literature. These payments were significantly larger than previous economic stimulus payments, summing to a maximum of \$11,400 for a family of four over a period of less than one year. The EIPs were also distributed more quickly than similar programs in the past; as a result, they were not staggered by social security number and thus identification of their effects is not as clean as was possible in 2001 and 2008. Furthermore, they were distributed at a time of highly unusual consumption behavior, both at the aggregate level and between income groups and geographies, all of which make it harder to tease out their spending effects. Nevertheless, the EIPs have proved a fruitful source of knowledge about the effectiveness of stimulus programs. 

\citet{parker2022EIP} is closely aligned with the methodology of \citet{parker2006rebates} and \citet{parker2013stimulus}, making use of the CEX, but, because the randomly staggered research design is not available, this newer analysis ``leans on comparing the spending of similar households that do and do not receive EIP and that receive EIPs of different amounts relative to their typical spending amounts''. They find that households spent their EIPs more slowly on average than the stimulus payments in 2001 and 2008.

Relative to the 2001 and 2008 stimulus payment studies, by 2020 many researchers had access to transaction level data. These data mitigate concerns about survey respondents under reporting their spending but can suffer more from sample selection bias than surveys designed by statistical agencies, such as the CEX. \citet{karger2020heterogeneity}, \citet{misra2022impact}, and \citet{baker2023income} all make use of data from personal finance apps. These convenience samples skew somewhat towards low-income households, while the SaverLife app studied in \citet{baker2023income} is specifically aimed at helping users to save money. All these studies find relatively high MPCs within the first few weeks after the first EIP payment is received---somewhere between 0.25 and 0.5 on average. Furthermore, and in contrast to some other studies, \citet{baker2023income} finds little response of durables spending. 

Recently, there has been a growing interest in heterogeneous agent models in macroeconomics that has increased the demand for reliable estimates of MPC heterogeneity, with a particular focus of the role of liquidity in determining MPCs. Some of the best evidence on MPC heterogeneity has come from outside of the United States. In \citet{fagereng2019lottery}, the authors study the spending response of Norwegian lottery winners using administrative data. They find that, in line with buffer-stock models, MPCs are negatively correlated with the winner's stock of liquid assets and negatively correlated with the size of the lottery win. They also find that MPCs decline with age, even controlling for liquid assets. Despite these first two correlations going in the expected directions, the high level of spending in the year in which the lottery is won, especially for households with significant liquid asset holdings, is difficult to reconcile with standard models. A further benefit of this study of Norwegian lottery winners is that the consumption response can be tracked over several years and is thus informative not just of the initial MPC but can also be used to inform the so-called intertemporal MPC---how households react to a shock to their income over time. The evidence suggests that spending is more front-loaded than a standard model would predict, especially given the large size of the lottery wins studied. 

In a study of French households, \citet{boehm2023fivefacts} conduct a randomized controlled trial to analyze the consumption response to unanticipated one-time money transfers of \euro300. The trial randomized three different types of transfer via a prepaid debit card: one transfer with no restrictions, one transfer in which unspent value expired within three weeks, and one transfer subject to a 10\% negative interest rate every week. The authors were able to see participants main bank accounts and therefore the effect of the transfers on their overall spending. They find an overall MPC of 0.23 within the first month of the transfer. Spending is concentrated in the first few weeks following the transfer, after which there is little boost to spending. They observe significant MPC heterogeneity by observable household characteristics including by liquid wealth, current and permanent income, and gender. However, similar to the Norwegian lottery study, they find that even households with high liquid asset holdings have high MPCs and---something that is not observable in the Norwegian data---that the spending response is concentrated in the short run for non-durables. Finally, MPCs are highest for the group given the card that becomes unusable after three weeks and lowest for the group given the card that has no restrictions on its use. Many other studies have found correlations between MPCs and observables, particularly a negative relation between MPCs and liquid wealth. However, observables are likely to only explain a small fraction of all MPC heterogeneity, as documented by \citet{lewis2021heterogeneity}.

\begin{table}[t!]
\begin{center}
\caption{Summary of Estimates from Natural Experiments and Elicitation Surveys}\label{Table::SummaryOther}
\resizebox{\textwidth}{!}{%
\begin{tabular}{l C{4cm} C{2.5cm} C{1.0cm} L{4.5cm}}
\toprule
Natural experiment study           &\textbf{MPC}                & Horizon           & $C_{it}$      & Data          \\
\midrule
\citet{baker2023income}            & 0.25$-$0.40                & First weeks       & te            & SaverLife 2020 \\
\citet{boehm2023fivefacts}         & 0.23                       & One Month         & te            & French RCT 2022\\
\citet{fagereng2019lottery}        & 0.35$-$0.71                & First year        & te            & N 1993-2015 \\
\citet{gelman2023gas}              & $\approx$1.00              & Perm. shock       & te            & FA 2013-16  \\
\citet{parker2006rebates}          & 0.2$-$0.40                 & Three months      & nde           & C 2001\\
\citet{MisraSurico2014_EvidenceFiscalStimulus}  & 0.43          & Three months      & nde           & C 2001\\
                                   & 0.16                       & Three months      & te            & C 2008\\
\citet{karger2020heterogeneity}    & 0.46                       & Two weeks         & te            & Facteus 2020\\
\citet{misra2022impact}            & 0.29                       & A few days        & te            & Facteus 2020\\
\citet{orchard2023counterfactuals} & $\approx$ 0.3              & Three months      & te            & C 2008\\
\citet{parker2022EIP}              & 0.05$-$0.16                & Three months      & nde           & C 2020-21\\
\citet{parker2013stimulus}         & 0.50$-$0.90                & Three months      & te            & C 2008\\
\citet{sahm2010household}          & $\approx$0.3               & One year          & te            & C 2008\\
\midrule
Elicitation survey study           &\textbf{MPC}                & Horizon           & $C_{it}$      & Data         \\
\midrule
\citet{bunn2018asymmetry}          & 0.14 (pos.) 0.64 (neg.)    & One year          & te            & BoE survey 2011-14\\
\citet{christelis2019asymmetric}   & 0.20 (pos.) 0.24 (neg.)    & One year          & te            & Dutch survey 2015\\
\citet{colarieti2024howandwhy}     & 0.16                       & One quarter       & te            & Authors' survey 2022-23\\
\citet{fuster2020what}             & 0.07 (pos.) 0.32 (neg.)    & Three months      & te            & NY Fed SCE 2016-17\\
\citet{jappelli2014fiscal}         & 0.48                       & Unspecified       & te            & SHIW 2010\\
\citet{jappelli2020reportedMPC}    & 0.47                       & Unspecified       & te            & SHIW 2016\\
\bottomrule
\end{tabular}}
\caption*{\fsz\emph{Legend:} \textbf{neg.}: negative; \textbf{nde}: non-durable expenditure; \textbf{pos.}: positive; \textbf{te}: total expenditure; \textbf{BoE}: Bank of England; \textbf{C}: Consumer Expenditure Survey; \textbf{FA} Financial Aggregator; \textbf{N} Norwegian registry data; \textbf{SCE}: Survey of Consumer Expectations; \textbf{SHIW}: Italian Survey of Household Income and Wealth.}
\end{center}
\end{table}

A standard income shock like those above mixes both the effects of an increase in lifetime budget constraint with the effects of an increase in liquidity. \citet{hamilton2023earlypension} analyze a policy implemented in Australia during the pandemic in which individuals were able to withdraw up to A\$20,000 from their government retirement accounts thus increasing liquidity without a change to individual's lifetime budget constraint. The authors find about one in six eligible people withdrew, and, furthermore, of those who did, most withdrew the maximum amount possible and spent close to half in the first four weeks. 

A smaller and negative liquidity shock is studied in \citet{gelman2020shutdown}. The authors look at the period during which the U.S. government shut down in 2013 and many federal workers received no (or reduced) pay which was made up to them later. Large negative spending effects---implying an MPC close to 0.5---were observed. However, much of this effect can be classified as accessing non-standard sources of short-term liquidity, in particular delaying recurring payments such as mortgage and rent payments.

Most natural experiments in this literature, including all those cited thus far, have been related to transitory income shocks. Less is known from the natural experiment literature about the consumption response to permanent shocks to income. \citet{gelman2023gas} use changes in gas prices as a proxy for permanent income changes and find, using transaction-level data, that individuals reduce their spending one-for-one with a permanent change in their disposable income. \citet{gerard2021jobdisplacement} look at displaced workers in Brazil who receive a positive transitory shock to income in the form of severance pay, but face a permanent reduction in their lifetime earnings. The authors find that workers increase spending at layoff by 35 percent despite experiencing a 14 percent long-term loss.

While this overview of the natural experiment literature has focused on income \emph{shocks}, there is also a large literature examining the consumption response to \emph{anticipated} income changes. For example, \citet{ganong2019unemployment} look at spending around the expiration of unemployment benefits, \citet{hsieh2003alaska} and \citet{kueng2018alaska} look at spending from the Alaska permanent fund, and \citet{souleles1999taxrefunds} and \citet{gelman2022rational} consider spending around tax refunds. Although standard consumption theory would suggest households' response to anticipated changes should be significantly muted relative to shocks, many of these papers show a large spending change coinciding with the anticipated income change.

\section{Elicitation Surveys}\label{Section::ElicitationSurveys}

The third method that researchers use to understand households' responses to income shocks is to ask them directly in a survey. This has the advantage of being relatively easy to implement without the need to find a natural experiment or get access to data on spending. The method also allows for direct comparison within a household of how their consumption would change in a variety of scenarios, such as in response to permanent versus transitory shocks, or to shocks of different size and sign. However, survey respondents may be unable to know their true behavior. 

\citet{shapiro1995incometiming,shapiro2003consumer} started this literature, asking households about their qualitative response to income changes. \citet{jappelli2014fiscal,jappelli2020reportedMPC} explore heterogeneity, particularly across cash-on-hand holding, using the Italian Survey of Household Income and Wealth which asks respondents for numerical MPCs. \citet{bunn2018asymmetry} and \citet{christelis2019asymmetric} both explore asymmetry in consumption responses, finding that responses to losses are larger than to gains. \citet{fuster2020what} ask households what they would do with \$500 in the NY Fed survey on consumer expectations. This survey also asked participants about their responses to news about future income shocks and found that even those with large responses to gains do not respond to news about future gains, although they do cut spending in response to news about future losses. \citet{colarieti2024howandwhy} further explore the dynamic spending response to news about future income shocks by asking for spending plans over four quarters. These authors also carry out several cross-validation exercises which suggest that the answers elicited from this type of survey align with actual spending behavior, suggesting these survey methods are of high value.

\section{Conclusion}\label{Section::Conclusion}

We survey the economics literature of, primarily, the last 20 years, that studies how household consumption responds to income shocks. We group the papers in this literature into three categories: papers that use structural methods, those that exploit natural experiments, and papers that rely on elicitation surveys. The evidence so far suggests that consumption responds more strongly to permanent than to transitory shocks, the sign and size of shocks as well as the horizon over which effects are studied matter, and there is substantial heterogeneity in the responses that cannot be fully explained by observable factors. This is an increasingly busy literature that will remain active in the foreseeable future.


\begin{singlespace}
\bibliographystyle{chicago}
\bibliography{consumptionencyclopedia.bib}
\addcontentsline{toc}{section}{Bibliography} 
\end{singlespace}
\end{document}